\documentclass[aip, preprint, groupedaddresses,superscriptaddress,nofootinbib]{revtex4-1} 

\usepackage[authormarkuptext=name,authormarkup=none]{changes}
\definecolor{acolour}{RGB}{0, 0, 255}
\definecolor{red}{RGB}{255, 0, 0}
\definecolor{green}{RGB}{0, 192, 0}
\definechangesauthor[name={Yorick}, color={acolour}]{YS}
\definechangesauthor[name={Gianluca}, color={red}]{GL}
\definechangesauthor[name={Alec}, color={green}]{AES}

\usepackage{graphicx}
\usepackage{dcolumn}
\usepackage{bm}
\usepackage{amsmath}
\usepackage[T1]{fontenc}
\usepackage[utf8]{inputenc}
\usepackage{etoolbox}
\usepackage{hyperref}
\usepackage{braket}

\newcommand*{\matr}[1]{\mathbf{#1}}
\newcommand*{\vect}[1]{\bm{#1}}

\makeatletter
\def\@email#1#2{%
 \endgroup
 \patchcmd{\titleblock@produce}
  {\frontmatter@RRAPformat}
  {\frontmatter@RRAPformat{\produce@RRAP{*#1\href{mailto:#2}{#2}}}\frontmatter@RRAPformat}
  {}{}
}%
\makeatother

\begin{document}

\preprint{AIP/123-QED}

\title{
Orbital-optimized Density Functional Calculations of Molecular Rydberg Excited States with Real Space Grid Representation and Self-Interaction Correction
}

\author{Alec E. Sigurðarson}
\affiliation{Science Institute and Faculty of Physical Sciences, University of Iceland, Reykjavík, Iceland}
\author{Yorick L. A. Schmerwitz}
\affiliation{Science Institute and Faculty of Physical Sciences, University of Iceland, Reykjavík, Iceland}
\author{Dagrún K. V. Tveiten}
\affiliation{Science Institute and Faculty of Physical Sciences, University of Iceland, Reykjavík, Iceland}
\author{Gianluca Levi}
\affiliation{Science Institute and Faculty of Physical Sciences, University of Iceland, Reykjavík, Iceland}
\author{Hannes Jónsson} 
\affiliation{Science Institute and Faculty of Physical Sciences, University of Iceland, Reykjavík, Iceland}
\affiliation{Dept. of Chemistry, Brown University, Providence, RI, USA}
\email{hj@hi.is, hannes$\_$jonsson@brown.edu}

\begin{abstract}
Density functional calculations of Rydberg excited states up to high energy are carried out for several molecules using an approach where the orbitals are variationally optimized by converging on saddle points on the electronic energy surface within a real space grid representation. Remarkably good agreement with experimental estimates of the excitation energy is obtained using the generalized gradient approximation (GGA) functional of Perdew, Burke and Ernzerhof (PBE) when Perdew-Zunger self-interaction correction is applied in combination with complex-valued orbitals.
Even without the correction, the PBE functional gives quite good results despite the fact that corresponding Rydberg virtual orbitals have positive energy in the ground state calculation. 
Results obtained using the TPSS and r2SCAN meta-GGA functionals are also presented, but they do not provide a systematic improvement over the results from the uncorrected PBE functional.
The grid representation combined with the projector augmented-wave approach gives a simpler and better representation of the diffuse Rydberg orbitals than a linear combination of atomic orbitals with commonly used basis sets, the latter leading to an overestimation of the excitation energy due to confinement of the excited states.
\end{abstract}

\maketitle


\section{\label{sec:Intro} Introduction}


The calculation of excited electronic states is growing in importance, as fields such as ultrafast spectroscopy, solar energy conversion, and photocatalysis are in rapid development. There, it can be important to model the dynamics of the atoms after excitation. Such simulations require a method for describing excited states that combines computational efficiency with sufficient accuracy and ability to provide forces acting on the atoms in the excited state. 
High-level wave function based approaches such as multireference configuration interaction and coupled cluster methods are known to give accurate results, but are typically too computationally intensive for general use, especially for large systems, as the computational effort scales rapidly with system size.

Density functional theory (DFT) and Kohn-Sham (KS) functionals have been developed and found to be highly successful for the description of electronic systems in the ground state, even for large systems, as the computational effort scales relatively slowly with size.\cite{Kohn1965, Hohenberg1964} The Kohn-Sham equations are solved to obtain orbitals from which a single determinant approximation to the ground electronic state is constructed. While an elaborate formalism has been developed to extend DFT to excited states by introducing time dependence in so-called time-dependent DFT (TD-DFT),\cite{Runge1984} there is increasing interest in exploring a simpler approach where higher-energy solutions of the KS equations are obtained.\cite{Perdew1985,Gorling1999,Ayers2012,Ayers2015,Hait2021, Levi2020jctc,Giarrusso2023} The practical problem faced by such time-independent calculations of excited states is collapse to a lower-energy solution than the target excited state or even to the ground state in the self-consistent field (SCF) iterations. This problem is addressed in the so-called maximum overlap method (MOM)\cite{Gilbert2008,Barca2018} by constraining the SCF iterations to maintain maximum overlap with an initial guess of the excited state determinant. However, MOM has been found to be unable to prevent variational collapse in a number of cases, one type of examples being intramolecular charge transfer excited states.\cite{Hait2020jctc, Carter-Fenk2020, Ivanov2021, Levi2020jctc,Schmerwitz2022,Schmerwitz2023} An alternative approach involves minimizing the squared energy gradient or the energy variance,\cite{Hait2020jctc, Zhao2019} but then convergence can be problematic due to the presence of unphysical stationary points on the variance optimization landscape.\cite{Burton2022, Cuzzocrea2020, Carter-Fenk2020} Another method for reducing the risk of variational collapse without using MOM is to raise the energy of the unoccupied orbitals in a so-called level shifting procedure, but this approach tends to reduce the size of the optimization steps\cite{Carter-Fenk2020} and increase the number of iterations required to reach convergence. More recently, methods have been developed where the stationary, i.e.\,variationally optimized, excited state solutions of the KS equations are obtained using 
algorithms specifically designed to converge on saddle points of the electronic energy surface, thereby avoiding collapse to a lower energy state.\cite{Schmerwitz2023, Ivanov2021, Levi2020jctc, Levi2020fd} 

In time-independent density functional calculations of excited states, the atomic forces can be evaluated easily since a stationary solution to the KS equations is obtained and the Hellman-Feynman theorem holds. Moreover, this approach has been found to give better results than commonly used, practical approximations to the TD-DFT formalism, such as the linear-response approximation,\cite{Casida1995} especially when going beyond the lowest excited state,\cite{Hirao2023, Besley2021, Hait2020jpcl, Seidu2014, Besley2009, Cheng2008} and when a significant rearrangement of the electron distribution occurs, such as in charge transfer and doubly excited states.\cite{Hait2020jctc, Carter-Fenk2020, Barca2018, Hait2016, Zhekova2014, Dreuw2004} Furthermore, the description of potential energy surfaces around conical intersections, where electronic states are near-degenerate, is problematic in typical TD-DFT calculations.\cite{Levine2006} The time-independent calculations can give better results in such cases\cite{Schmerwitz2022, Maurer2011} and moreover, require only slightly greater computational effort than ground state calculations. For example, it has recently been shown that an avoided crossing and conical intersection in the ethylene molecule are accurately described provided the symmetry of the excited state wave function is allowed to break.\cite{Schmerwitz2022} Lastly, time-independent density functional calculations have also successfully been applied in simulations of the dynamics of organic molecules and transition metal complexes, even incorporating the interaction with explicitly included solvent molecules.\cite{Vandaele2022, Levi2020pccp, Levi2018, Pradhan2018}

Excited electronic states, even the lowest excited state for some molecules, can correspond to highly diffuse orbitals. The energy of successively higher states of this type approximately follow the Rydberg formula and are referred to as Rydberg states.
Due to the self-interaction error inherent in practical, semi-local implementations of KS functionals, the long-range form of the effective potential for an electron does not have the correct -1/r form, unless it is explicitly enforced in some way, so the applicability of such functionals in calculations of Rydberg states could be called into question. The virtual orbitals that could correspond to Rydberg excited electrons typically 
have positive orbital energy in ground state calculations using KS functionals
and thereby represent unbound electrons. If, however, an explicit Perdew-Zunger self-interaction correction\cite{Perdew1981} (PZ-SIC) is applied, the long-range interaction naturally approaches -1/r and bound diffuse orbitals are obtained in the ground state calculation. The same issue affects ground state calculations of weakly bound electrons in, for example, anions that can turn out to be unstable in semi-local KS functional calculations. An example of this effect is a dipole bound anion where the Perdew-Burke-Enzerhof (PBE) functional gives a positive long-range potential and the anion is predicted to be unstable, while PBE-SIC correctly gives a potential corresponding to a point dipole at long range and predicts a binding energy in close agreement with the experimental value.\cite{zhang_2016}  PZ-SIC has been used in non-self-consistent calculations of Rydberg states of various molecules and even molecular clusters, where an electron is placed in a frozen ground state Rydberg orbital to describe the excited electron while the other orbitals are optimized 
in a DFT calculation\cite{Gudmundsdottir2013,Zhang2017,Gudmundsdottir2014,Cheng2016}. Alternatively, self-consistent calculations of Rydberg excited states using KS functionals without any correction, even the local density approximation, have given surprisingly good results for small molecules and atoms.\cite{Seidu2014, Cheng2008}


The primary question addressed in this article is how well fully variational, self-consistent calculations using a self-interaction corrected semi-local KS functional can predict the excitation energy to Rydberg singlet and triplet states of molecules. To this end, orbital-optimized calculations are carried out for a total of 
31
singlet and triplet Rydberg excitations in four molecules, C\textsubscript{2}H\textsubscript{4}, CH\textsubscript{2}O, NH\textsubscript{3} and H\textsubscript{2}O, ranging up to 10\,eV above the ground state. Calculations using the PBE functional\cite{Perdew1996}, which is of generalized gradient approximation (GGA) form, are compared with calculations using PZ-SIC applied to PBE as well as calculations using two meta-GGA functionals that include higher-order derivatives, the one by Tao, Perdew, Staroverov and Scuseria (TPSS)\cite{Perdew2004} and the more recent r2SCAN functional.\cite{sun_strongly_2015,Furness2020} The calculations are performed using a recently developed direct orbital optimization method for excited states\cite{Ivanov2021} where the orbitals are represented 
on a real space grid in combination with the projector augmented-wave approach (PAW).\cite{Blochl1994}
Such a representation can more accurately describe diffuse Rydberg orbitals than more commonly used linear combination of atomic orbitals (LCAO) basis sets that require the addition of specifically designed diffuse functions. The calculated excitation energy values are compared to experimental estimates as well as published results obtained from high-level quantum chemistry calculations\cite{Loos2018,Feller2014,Li2006}.

The article is structured as follows. First, a discussion of the theory underlying the method is presented and then a summary of the computational settings. The results of the calculations of the excitation energy values are then presented for each molecule and comparison made with the experimental estimates as well as theoretical estimates coming from published results of calculations using configuration interaction and coupled cluster methods. The article ends with a discussion and a conclusion section.


\section{\label{sec:Methods} Methodology}

\subsection{Excited Electronic States by Direct Optimisation}

Excited states of electronic systems can be described as solutions of the KS equations with non-aufbau occupation of the orbitals\cite{Perdew1985,Gorling1999,Ayers2012,Ayers2015,Hait2021,Giarrusso2023}. These solutions correspond to stationary points other than the ground state minimum on the surface describing the variation of the energy of the system as a function of the electronic degrees of freedom. These stationary points typically correspond to saddle points since the energy of the system decreases as one or more of its degrees of freedom are altered toward the ground state. 

The orbitals $\vect{\psi} = \left(\ket{\psi_1}, \ldots, \ket{\psi_{M}}\right)^{\mathrm T}$ can be expanded in a linear combination of auxiliary orbitals $\vect{\phi} = \left(\ket{\phi_1}, \ldots, \ket{\phi_{M}}\right)^{\mathrm T}$, which can be the occupied and virtual orbitals of the ground state, or any other initial guess orbitals for the excited state calculation
\begin{equation}
 \vect{\psi} = \matr{U}\vect{\phi}\,.
\end{equation}
The energy is a functional of both $\vect{\phi}$ and the $M\times M$ unitary matrix $\matr{U}$. The stationary states can be obtained through an iterative double loop procedure, where at each iteration a stationary point in the space of unitary matrices $\matr{U}$ is found, followed by a step of minimization with respect to $\vect{\phi}$
\begin{equation}
 \underset{\vect{\psi}} {\mathrm{stat}\,} E_{\mathrm{KS}}[\vect{\psi}] = \underset{\vect{\phi}} {\mathrm{min}\,}
 \underset{\matr{U}} {\mathrm{stat}\,} E_{\mathrm{KS}}[{\matr{U}}\vect{\phi}]\,.
\end{equation}
In the direct optimization (DO) method developed for plane wave and grid representations of the orbitals\cite{Ivanov2021} and used in the present work, $\matr{U}$ is parameterized through the exponential transformation $\matr{U} = e^{\boldsymbol{\kappa}}$, where
\begin{equation}
\boldsymbol{\kappa} = \begin{pmatrix}\boldsymbol{\kappa}_{\mathrm{oo}} & \boldsymbol{\kappa}_{\mathrm{ov}}\\ -\boldsymbol{\kappa}_{\mathrm{ov}}^{\dag} & \boldsymbol{\kappa}_{\mathrm{vv}}\end{pmatrix}
\end{equation}
is an anti-Hermitian matrix ($\boldsymbol{\kappa} = -{\boldsymbol{\kappa}^\dagger}$) containing blocks of orbital rotations between occupied-occupied (oo), occupied-virtual (ov) and virtual-virtual (vv) orbitals. The energy is always invariant to ${\boldsymbol{\kappa}}_{\mathrm{vv}}$ and, for functionals that do not depend on the orbital densities, also invariant to orbital rotations among equally occupied orbitals. Convergence on stationary points corresponding to excited states can then be obtained using efficient quasi-Newton algorithms for finding saddle points in the linear space of anti-Hermitian matrices\cite{Levi2020jctc, Levi2020fd} together with an outer loop minimization following the gradient projected on the tangent space at $\vect{\phi}$.

In the LCAO representation, the coefficients $\matr{C}_{0}$ corresponding to the auxiliary orbitals are fixed. Therefore, only an optimization with respect to the elements of $\boldsymbol{\kappa}$ is carried out to find the optimal orbital coefficients $\matr{C} = \matr{C}_{0}e^{\boldsymbol{\kappa}}$. The LCAO calculations presented here employ a recently developed generalized mode following method.\cite{Schmerwitz2023} There, excited states are found as saddle points of a given order $n$ by inverting the electronic gradient along the modes corresponding to the lowest $n$ eigenvalues of the electronic Hessian, thereby avoiding collapse to lower energy states without any use of MOM.


\subsection{Self-Interaction Correction}

The Hartree energy term in KS functionals 
contributes
a potential for each electron that is formed by all the electrons, 
including the electron 
considered. This self-interaction is incorrect, as electrons should not be affected by a potential formed by themselves. 
Even in the case of a system with only one electron,
this spurious interaction leads to  
non-zero energy due to the electron interacting with itself. This self-interaction error should be corrected by the exchange-and-correlation (XC) part of the functional, but the semi-local form of practical, 
approximate KS (AKS) 
functionals renders them unable to compensate fully since the self-interaction is non-local. 
An explicit non-local correction can, however, be applied orbital-by-orbital as proposed by Perdew and Zunger (PZ-SIC, below simply referred to as SIC).\cite{Perdew1981} The self-interaction corrected functional can then be expressed as
\begin{equation}
    E_{\mathrm{SIC}}\left[
\rho_{1}, \rho_{2} \dots \rho_{N}
\right] = E_{\mathrm{AKS}}\left[\rho\right] - \sum_{i}^{N} \left(E_{\mathrm{H}}\left[\rho_{i}\right] + E_{\mathrm{XC}}\left[\rho_{i}\right]\right)\,,
\end{equation}
where $E_{\mathrm{AKS}}$ is the 
energy given by an approximate functional of the KS form,
$E_{\mathrm{H}}$ the Hartree energy, $E_{\mathrm{XC}}$ the exchange-correlation 
energy, and $\rho$ and $\rho_i$ are the total electron density and electron density associated with orbital $i$, respectively.

The application of PZ-SIC makes the
functional orbital density dependent 
and the energy is, therefore, no longer invariant with respect to a unitary transformation of the occupied orbitals. As such, finding stationary states of $E_{\mathrm{SIC}}$ involves an additional unitary optimization among the occupied orbitals to maximize the term 
$\sum_{i}^{N}\left(E_H\left[\rho_{i}\right] + E_{\mathrm{XC}}\left[\rho_{i}\right]\right) .$
\cite{Lehtola2016, Klupfel2011, Klupfel2012} 
When DO is used with SIC, this maximization can be accomplished though an additional inner loop minimization\cite{Ivanov2021}
\begin{equation}
 \underset{\vect{\psi}} {\rm stat\,} E_{\mathrm{
 SIC
 }}
 = \underset{\vect{\phi}} {\rm min\,}
 \underset{\matr{U}} {\rm stat\,} \underset{\matr{U}_{\mathrm{oo}}} {\rm min\,} 
E_{\mathrm{ SIC}}[{\matr{U}}\vect{\phi}]\,,
\end{equation}
%
where $\matr{U}_{\mathrm{oo}}$ denotes the subspace of $\matr{U}$ corresponding to unitary transformation among the occupied orbitals.


Since this formulation of SIC calculations is variational, the Hellman-Feynman theorem applies, and atomic forces can be evaluated without significant additional computational effort. 
We note, however, that the application of SIC can in some cases break symmetry in the configuration of the atoms, as it leads to convergence on one out of two or more equivalent solutions with broken symmetry. One example is the benzene molecules where the relaxed PBE-SIC configuration has slight differences in C-C bond length, short and long bonds alternating.\cite{Lehtola2016}

PBE-SIC calculations of Rydberg states of molecules and molecular clusters have been presented earlier and used to help interpret experimental measurements,\cite{Gudmundsdottir2013,Zhang2017,Gudmundsdottir2014,Cheng2016}
but there the orbital corresponding to the excited electron has not been variationally optimized. The saddle point search method used here is fully variational and therefore, represents an improvement on the method used in those previous calculations.

In the PBE-SIC calculations presented here, no functional form is assumed for the orbitals, and a full variational optimization is carried out using complex-valued functions, thereby breaking unitary invariance and introducing orbital density dependence.  Other approaches have been taken in the implementations of PZ-SIC, such as the Fermi-Löwdin orbital formulation introduced to recast PZ-SIC into a unitary invariant form.\cite{Schwalbe2020,Pederson2023}

\subsection{Breaking Complex Conjugation Symmetry}
Point group symmetry can lead to degenerate orbitals and the exact excited state wave function then has multi-determinant character. In a KS calculation, when real-valued orbitals are used, as is commonly done, unequal occupation of the degenerate orbitals can lead to breaking of the spatial symmetry of the single determinant wave function. For example, the lowest excited state of the carbon monoxide molecules can be obtained by promotion of an electron from the HOMO to one of two degenerate $\pi^*$ orbitals ($\pi^*_x$ or $\pi^*_y$). The electron density of the resulting excited state solutions lacks uniaxial symmetry.\cite{Ivanov2021}
Multi-determinant effects may be captured even in a single-determinant formalism by choosing the orbitals to be complex-valued and forming complex linear combinations of the degenerate real-valued orbitals. The resulting single determinant wave function breaks complex conjugation and time reversal symmetries, but  
has an inherent multi-determinant character\cite{Perdew2021, Lee2019, Small2015}. Moreover, the electron density has the correct spatial symmetry.\cite{Ivanov2021, Lee2019, Small2015} This procedure has been shown to improve the value of the excitation energy for carbon monoxide.\cite{Ivanov2021}

\begin{figure}
    \centering
    \includegraphics[width=0.7\textwidth]{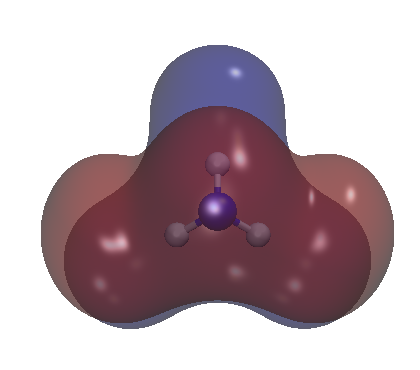}
    \caption{Comparison of the electron density of the spin-mixed HOMO$\rightarrow$LUMO+1 excited state of NH\textsubscript{3} when the excitation is to a real-valued $p_x$ or $p_y$ orbital (red), and when the excitation is to a complex-valued $p_+$ or $p_-$ orbital as in eq. \ref{eq:complex_orbs} (blue). In the first case, the density does not have the 3-fold rotation symmetry, while in the second case, the complex conjugation and time reversal symmetries are broken, but the density possesses the correct spatial symmetry. The rendering is for an isovalue of $0.001\,$\AA$^{-3}$.}
    \label{fig:density}
\end{figure}

Among the molecules calculated here, the electronic structure of  
NH\textsubscript{3} 
involves degenerate 
Rydberg $p_x$ and $p_y$ orbitals (the ground state LUMO+1). 
Assigning different occupation numbers to the two degenerate orbitals leads to an excited state solution with an electron density that lacks the 3-fold rotational symmetry of the molecule.
The spatial symmetry can be preserved by using complex-valued orbitals and breaking the complex conjugation and time reversal symmetries instead. 
Such an orbital set is created by forming a complex linear combination of the degenerate real-valued orbitals $p_x$ and $p_y$ obtained in a ground state calculation according to
\begin{align}\label{eq:complex_orbs}
    p_+ & = p_x + ip_y\,,\\
    p_- & = p_x - ip_y\,.
\end{align}
The complex-valued orbitals $p_+$ and $p_-$ break the complex conjugation symmetry since $\left(p_+\right)^{*} = p_-$ and the time reversal symmetry since $p_+$ and $p_-$ are unequally occupied. As illustrated in figure~\ref{fig:density}, unequal occupations of $p_x$ and $p_y$ break the spatial symmetry of the total density, while unequal occupation of $p_+$ and $p_-$ does not since $p_+$ and $p_-$ both have equal spatial contributions from $p_x$ and $p_y$. 
%

\subsection{Computational settings}
The density functional calculations are performed for molecules in vacuum and make use of the 
PBE, TPSS, and r2SCAN functionals with semi-local XC approximation as well as the orbital density dependent PBE-SIC functional. 
Four molecules are considered in this study: ethylene, formaldehyde, ammonia, and water. A total of 
31
excitations to Rydberg states are calculated. 
Experimentally determined molecular structures are used \cite{Haynes2016}.
For the ammonia and water molecules, even the first excited state is a Rydberg state, while for the ethylene and formaldehyde molecules, the first two are valence states while the higher ones are Rydberg states. 
 
The orbitals
are represented on a real space grid in combination with the PAW formalism,\cite{Blochl1994} except for the r2SCAN calculations.
The PBE-SIC calculations are carried out using complex-valued orbitals, as has been shown to be important even in ground state calculations of atoms and molecules.\cite{Klupfel2011,Klupfel2012} For comparison, Rydberg excitation energy obtained using real-valued orbitals in PBE-SIC calculations are also mentioned below to illustrate again the importance of using complex-valued orbitals with SIC. For all other functionals, real-valued orbitals are used except for the LUMO+1$\leftarrow$HOMO excited state of ammonia, where complex-valued orbitals are used to maintain spatial symmetry of the electron density (see previous section). The frozen core approximation is used within the PAW formalism.\cite{Blochl1994} The initial guess for the orbitals of an excited state is obtained from the ground state orbitals by choosing a non-aufbau occupation reflecting a single excitation of an electron from an occupied to a virtual orbital.

Both singlet and triplet Rydberg states are calculated. All singlet states are open-shell, and their energy is approximated using the spin-purification formula\cite{Ziegler1977}
\begin{equation}\label{eq:sum-rule}
    E_{S} = 2E_{M} - E_{T}\,,
\end{equation}
where $E_M$ is the energy of the spin-mixed determinant constructed by promotion of an electron within the same spin channel, while $E_T$ is the energy of the triplet state constructed by promotion of an electron between different spin channels. Both the spin-mixed and triplet states are variationally optimized independently before applying the spin-purification formula. 
The difference between the approximated singlet energy and the energy of the spin-mixed determinant is equal to
half of 
the resulting calculated singlet-triplet splitting energy of the state. The singlet-triplet splitting energy for the states considered here are in the range of 0.01\,eV to 0.40\,eV. The importance of using the spin purification formula is, in particular, evident in the calculations of the Rydberg states of H$_2$O where the ordering of the singlet $3p_z$ and $3p_x$ states is incorrect without the correction.

All calculations are performed with the GPAW software\cite{Mortensen2005, Enkovaara2010, Mortensen2023} and LIBXC\cite{Lehtola2018} version 5.1.6.
The calculations are converged to a precision of $10^{-6}\,\mathrm{eV}^{2}$ per valence electron in the squared residual of the KS equations
\begin{equation}
    \frac{1}{N}\sum_{i = 1}^{M}\int d\matr{r}f_{i}\left|\matr{\hat{H}}_{\mathrm{KS}}\psi_{i}\left(\matr{r}\right) - \sum_{j = 1}^{M}\lambda_{ij}\psi_{j}\left(\matr{r}\right)\right|^{2}\,,
        \label{eq:residual}
\end{equation}
where $\matr{\hat{H}}_{\mathrm{KS}}$ is the Kohn-Sham Hamiltonian operator and $f_i$ are the orbitals occupation numbers.
%
The unoccupied ground state orbitals are converged to a precision of 
$5 \times 10^{-4}$ in the norm of the gradient.




The calculations employ a cubic simulation cell with a minimum distance of 10\,\AA\, between the atomic nuclei and the edge of the cell.
The spacing between points in the uniform grid used to describe the valence electrons is 0.15\,\AA.
%
Tests were carried out to ensure that the size of the box is large enough and grid spacing small enough. For example, a calculation of the $3s$ state of ethylene using a simulation cell with a larger minimum distance of 11\,\AA\, instead of 10\,\AA\, gives excitation energy that differs by less than 0.2\,meV. A calculation with grid spacing of 0.10\,\AA\, instead of 0.15\,\AA\, gives excitation energy that differs by 1.5\,meV.

The r2SCAN calculations are instead performed using a plane wave expansion of the wave functions with a cutoff energy of 600\,eV.
The real space grid representation is 
apparently too coarse for this functional while the plane wave representation is smooth enough.
Calculations with r2SCAN are known to be hard to converge because of rapidly varying terms that require an exceptionally smooth representation of the orbitals. 

\begin{figure}
    \centering
    \includegraphics[width=\textwidth]{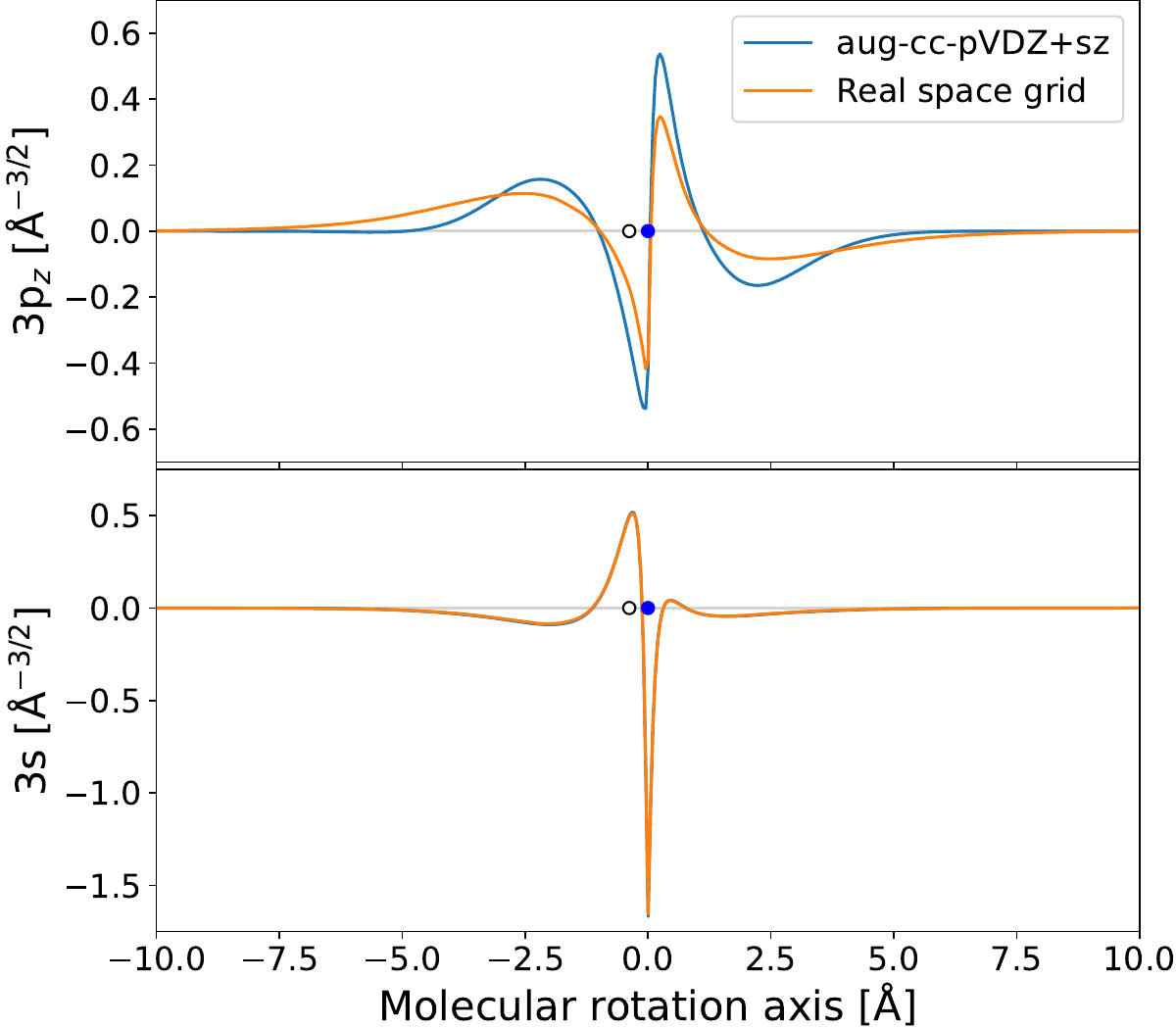}
    \caption{$3p_z$ and $3s$ Rydberg orbitals of ammonia obtained in excited state calculations using LCAO with an aug-cc-pVDZ+sz basis set compared to results of calculations using a real space grid representation with a grid spacing of 0.15\,\AA. 
    The PAW contribution is included.
    The two calculations are in reasonable agreement for the $3s$ state, while the $3p_z$ state shows clear limitations of the atomic basis set leading to large overestimation of the excitation energy as shown in figure 3. The position of the nitrogen nucleus along the $z$-axis is marked with a blue circle, while the white circle indicates the hydrogen atoms.
    }
    \label{fig:ammonia-orbitals}
\end{figure}

LCAO calculations are performed for comparison. There, valence electrons are represented by an atomic basis set consisting of primitive Gaussian functions taken from the aug-cc-pVDZ\cite{Dunning1989, Kendall1992, Woon1994} set augmented with a single set of numerical atomic orbitals (referred to as Gaussian basis set + sz)\cite{Rossi2015, Larsen2009} 
in combination with PAW. 
This basis set is comparable to the one used in 
some of
the high-level configuration interaction and coupled cluster calculations referred to below for comparison with the results of our density functional calculations.
It is challenging to represent Rydberg states with LCAO because of the wide distribution of the electron density, especially for the higher-energy states. This fact can, for example, be seen from calculations of the $3p_z$ Rydberg state of ammonia shown in figure~\ref{fig:ammonia-orbitals}. There is a clear difference between the long-range tails of the real space grid and LCAO results showing that the latter is incomplete and leads to confinement of the orbital.  The agreement is, however, quite good for the lower lying $3s$ Rydberg orbital, which is less diffuse.

\begin{figure}
    \centering
    \includegraphics[width=\textwidth]{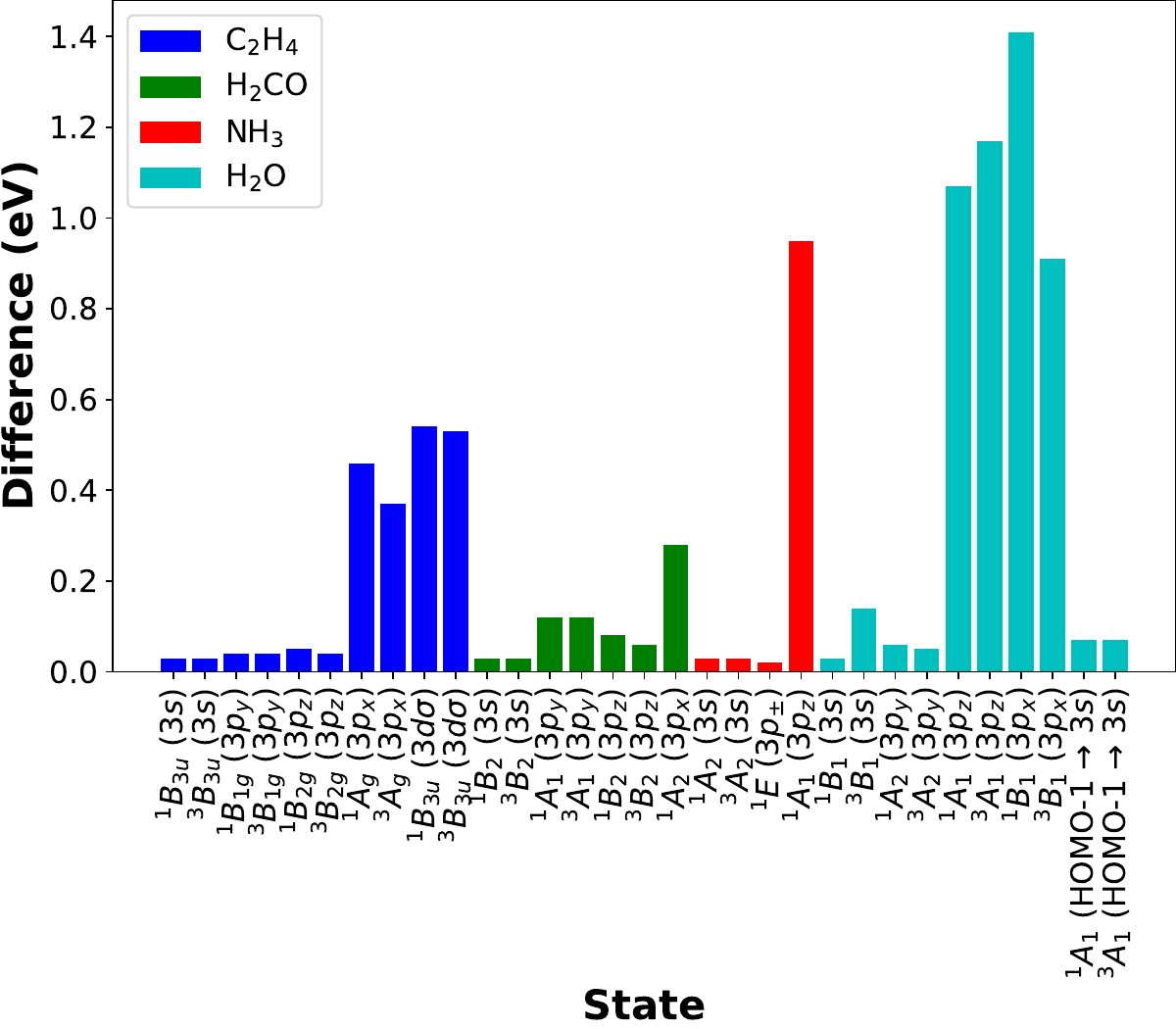}
    \caption{Lowering of the excitation energy for various Rydberg states of ethylene (blue), formaldehyde (green), ammonia (red), and water molecule (turquoise) when using a real space grid representation of the orbitals instead of the aug-cc-pVDZ+sz LCAO basis set.  
    The excitation energy is in all cases reduced because the excited state orbitals become less confined. The effect is most pronounced for excitation to the higher energy Rydberg states.}
    \label{fig:lcao-diff}
\end{figure}

Confinement of the Rydberg state orbitals in the LCAO calculations leads to an overestimation of the excitation energy as shown by the comparison of PBE calculations using the two basis sets in figure~\ref{fig:lcao-diff}. The overestimation of the excitation to the $3p_z$ Rydberg state of ammonia is particularly large, consistent with the lack of long-range tails illustrated in figure~\ref{fig:ammonia-orbitals}. For all 31 states calculated here, there is an overestimation of the excitation energy in the LCAO calculations with the basis set used here as compared to the grid based calculation, and this effect is greater for the higher-energy excitations.  
The LCAO basis set would need to include additional diffuse functions in order to better represent the Rydberg states.

\section{\label{sec:Results} Results}

\begin{figure}
    \centering
    \includegraphics[width=\textwidth]{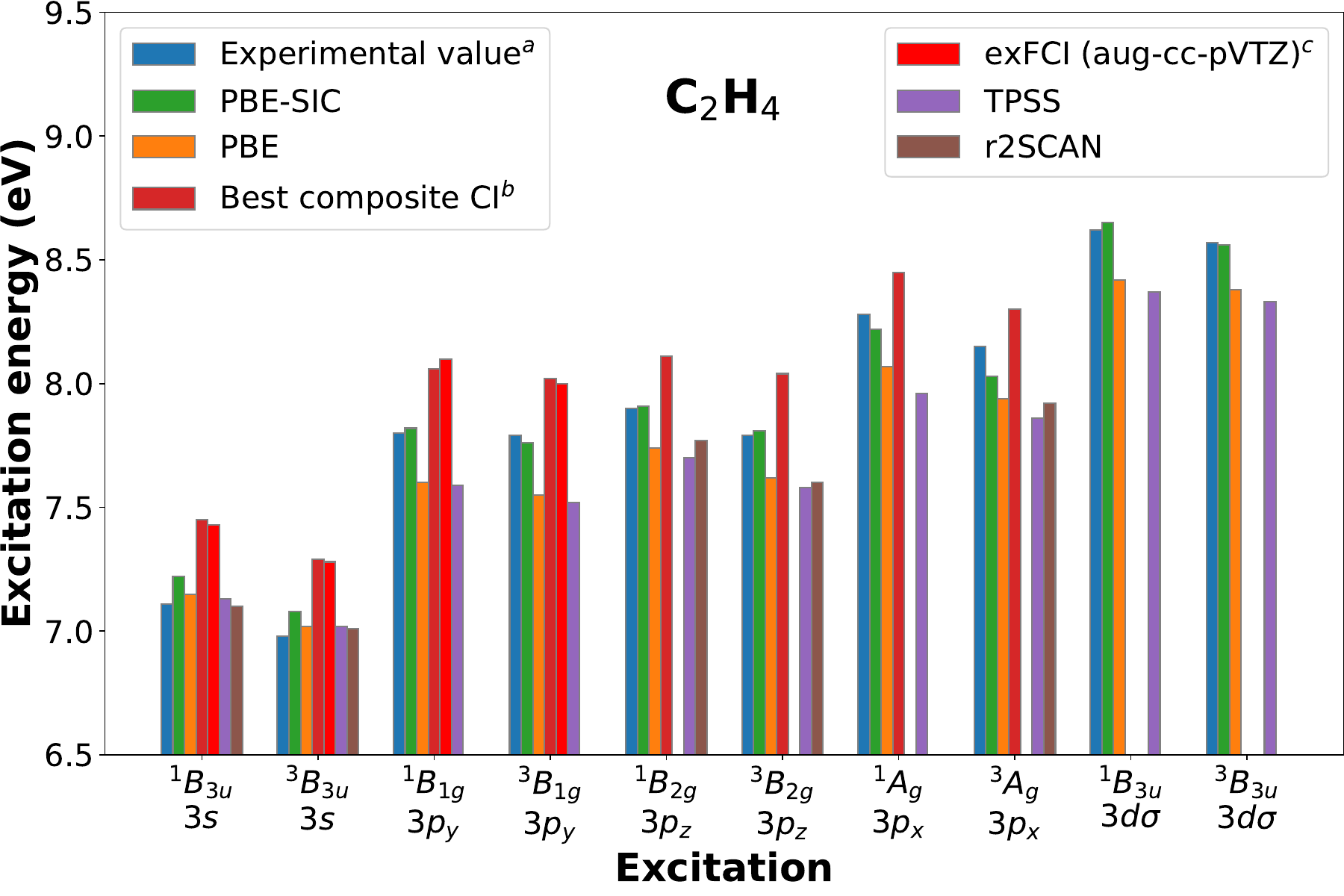}
    \caption{
    Vertical excitation energy for Rydberg states of ethylene estimated experimentally$^{a,}$\cite{Robin1985} (blue), and calculated with PBE-SIC (green) and PBE (orange). 
    Published results using a composite configuration interaction and coupled cluster method$^{b,}$\cite{Feller2014} (dark red) and an extrapolated full configuration interaction method$^{c,}$\cite{Loos2018} (light red), both with LCAO basis sets, are shown for comparison.
    Results using the meta-GGA functionals TPSS (purple) and, for some of the states, r2SCAN (brown) are also shown.
    While all the density functional calculations give quite good results for the excitations to the $3s$ Rydberg state, the GGA and meta-GGA calculations underestimate the excitation energy for the various $3p$ Rydberg states, while PBE-SIC gives values that are close to the experimental estimates. The configuration interaction and coupled cluster calculations give significantly higher excitation energy, possibly due to confinement effects resulting from an incomplete LCAO basis, as illustrated in figure 3.
    }
    \label{fig:ethylene}
\end{figure}

The results for ethylene are shown in figure \ref{fig:ethylene}. We have recently presented successful calculations of lower excited states, including a conical intersection between the ground and first valence excited state and an avoided crossing of the ground and lowest doubly excited state,\cite{Schmerwitz2022} but here we focus on the Rydberg excited states. A total of 10 excited states have been calculated, 5 singlet states and 5 triplet states.
The agreement between the experimental estimate from ref.~\citenum{Robin1985}
and the PBE-SIC calculation is remarkably good, even for the highest excitation energy of nearly 9\,eV. In most cases, the uncorrected PBE functional gives excitation energy values that are significantly lower than the experimental estimates. 

For the four lower-energy excitations, there are two published results from high-level quantum chemistry calculations: A composite configuration interaction and coupled cluster calculation \cite{Feller2014} and an extrapolated full configuration interaction calculation\cite{Loos2018}. Both make use of atomic basis sets in an LCAO representation of the orbitals. The two calculations agree well with each other, but are systematically giving higher excitation energy values than both the experimental estimates and the PBE-SIC calculations. This overestimation is possibly due to incompleteness of the atomic basis sets used. As shown in figure 3, a PBE calculation using 
a similar
atomic basis set, represented on the grid, indeed gives higher excitation energy than the PBE calculation using the real space grid. The diffuseness of the Rydberg state orbitals is hard to represent in atomic basis sets, as highly diffuse atomic functions need to be included. 

Results obtained using the TPSS meta-GGA functional are similar to the results obtained from PBE, systematically lower than the PBE-SIC results and experimental estimates, especially for the higher-energy excitations.
Calculations using the r2SCAN functional could only be made to converge for some of the excited states and give results that are similar to the TPSS values. Evidently, the additional complexity of the r2SCAN functional, which apparently is the reason for the difficulty in converging the calculations does not lead to a better description of the Rydberg states.

\begin{figure}
    \centering
    \includegraphics[width=\textwidth]{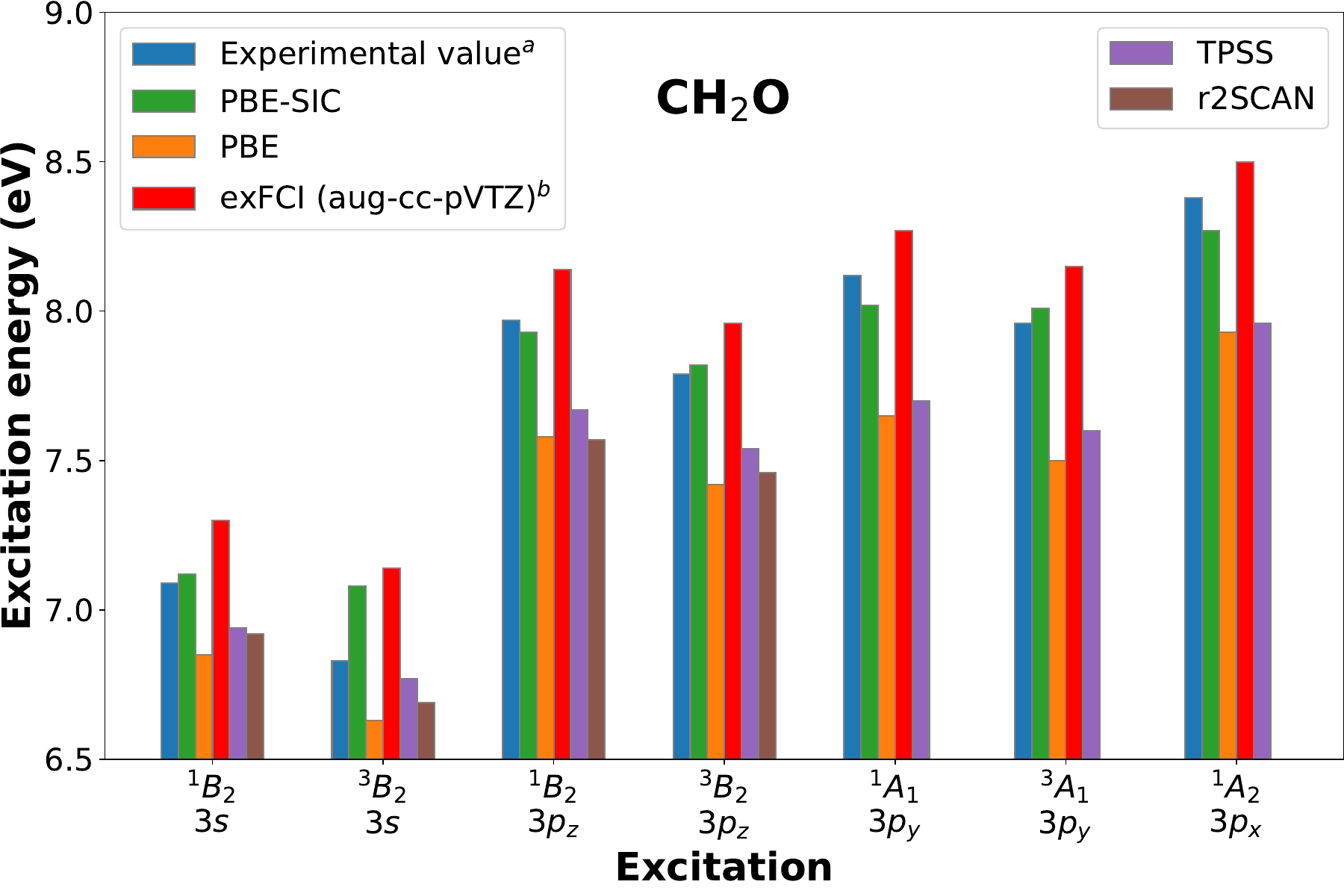}
    \caption{
    Vertical excitation energy for Rydberg states of formaldehyde estimated experimentally$^{a,}$\cite{Clouthier1983, Taylor1982} (blue), and calculated with PBE-SIC (green) and PBE (orange). 
    Published results of configuration interaction calculations$^{b,}$\cite{Loos2018} (light red) using LCAO are shown for comparison. They consistently yield higher excitation energy, possibly due to confinement resulting from incompleteness of the atomic basis set used there.
    The uncorrected PBE and meta-GGA functionals, TPSS (purple) and r2SCAN (brown), give significantly lower excitation energy. The r2SCAN calculations could only be made to converge for some of the excited states. 
    }
    \label{fig:formaldehyde}
\end{figure}

The results for formaldehyde are shown in figure~\ref{fig:formaldehyde}. A total of 7 excited states have been calculated, 4 singlet states and 3 triplet states. As for the ethylene molecule, the PBE-SIC results are close to the experimental estimates, except for the $3s$ triplet state. The uncorrected PBE and meta-GGA functionals, TPSS and r2SCAN, give significantly lower excitation energy (the r2SCAN calculations could only be converged for some of the excited states). Published results of quantum chemistry calculations, here using extrapolated configuration interaction method,\cite{Loos2018} are show substantially higher excitation energy, possibly due to incompleteness of the atomic basis set for the diffuse Rydberg states, as illustrated in figure 3. 

\begin{figure}
    \centering
    \includegraphics[width=\textwidth]{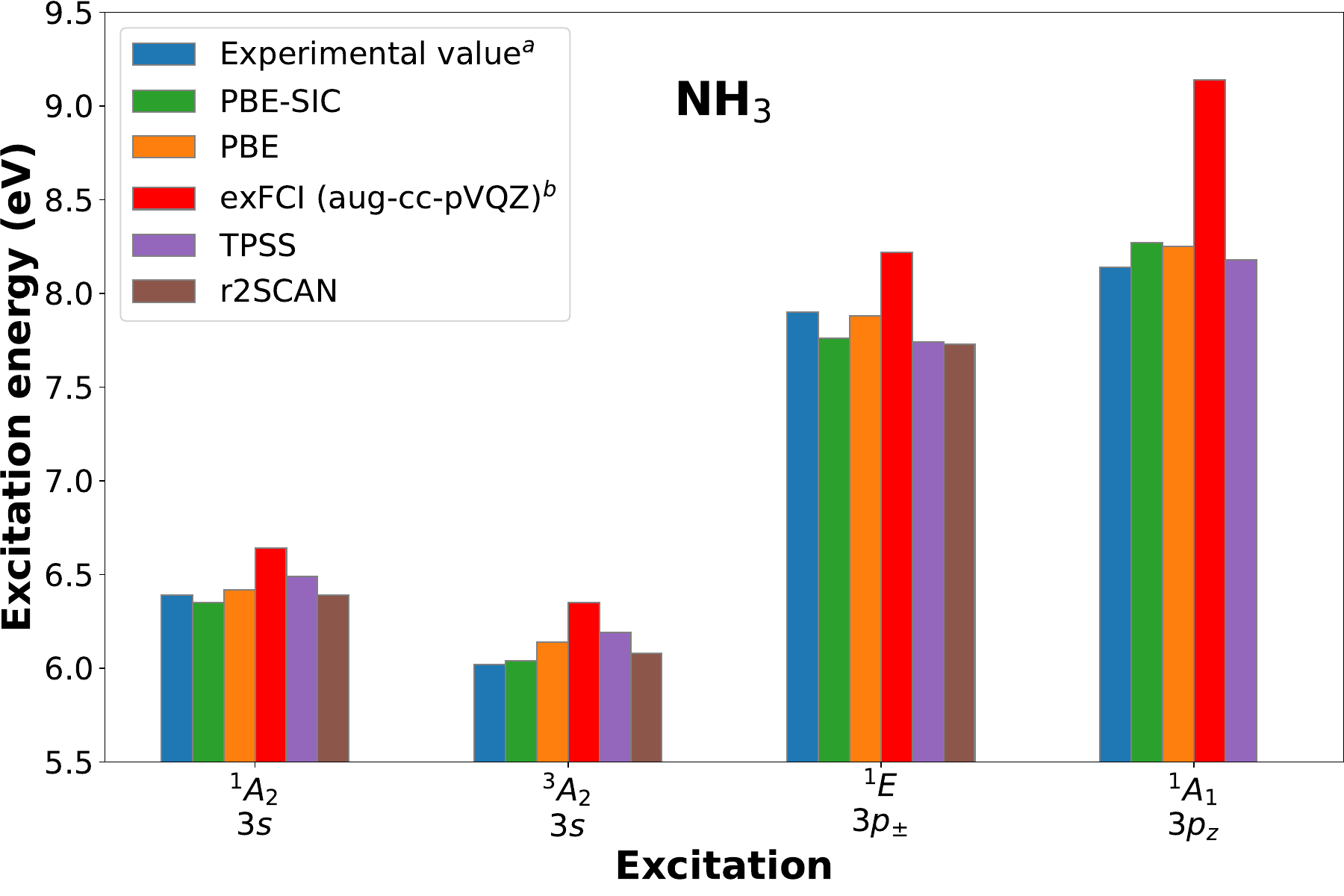}
    \caption{
    Vertical excitation energy for Rydberg states of ammonia estimated experimentally$^{a,}$\cite{Skerbele1965, Arfa_Tronc_1991} (blue), and calculated using PBE-SIC (green) and PBE (orange) functionals. Published results of configuration interaction calculations$^{b,}$\cite{Loos2018} (light red) using LCAO are shown for comparison. 
    The large value for the excitation energy for the $3p_z$ state is consistent with the large difference between grid based calculations and LCAO shown in figures 2 and 3.  
    Calculations using the meta-GGA functionals, TPSS (purple) and r2SCAN (brown), are also shown (the latter missing for the highest state because of problems with convergence). 
    The density functional calculations of the excited states make use of complex-valued orbitals (see section II.D).
}
    \label{fig:ammonia}
\end{figure}

The results for the ammonia molecule are shown in figure \ref{fig:ammonia}. The four Rydberg excited states for which experimental values are available have been calculated. Here, the results of the PBE-SIC calculations are quite similar to the results obtained with PBE and the meta-GGA calculations and are in fair agreement with the experimental estimates. For all four states, the published quantum chemistry calculations, based on extrapolated full configuration interaction method,\cite{Loos2018} give a significant overestimation of the excitation energy, especially for the highest state, $3p_z$. As illustrated in figures 2 and 3, this overestimation may be due to the incompleteness of the atomic basis set used in the LCAO calculations, while the grid based density functional calculations better represent the diffuse Rydberg orbitals. 

As discussed in section II.D, the calculations of the Rydberg excited state break the spatial symmetry of the electron density unless complex-valued orbitals are used, and the complex conjugation and time reversal symmetries are broken instead. When real-valued orbitals are used in the PBE calculation, the excitation energy turns out to be 0.1\,eV lower.

\begin{figure}
    \centering
    \includegraphics[width=\textwidth]{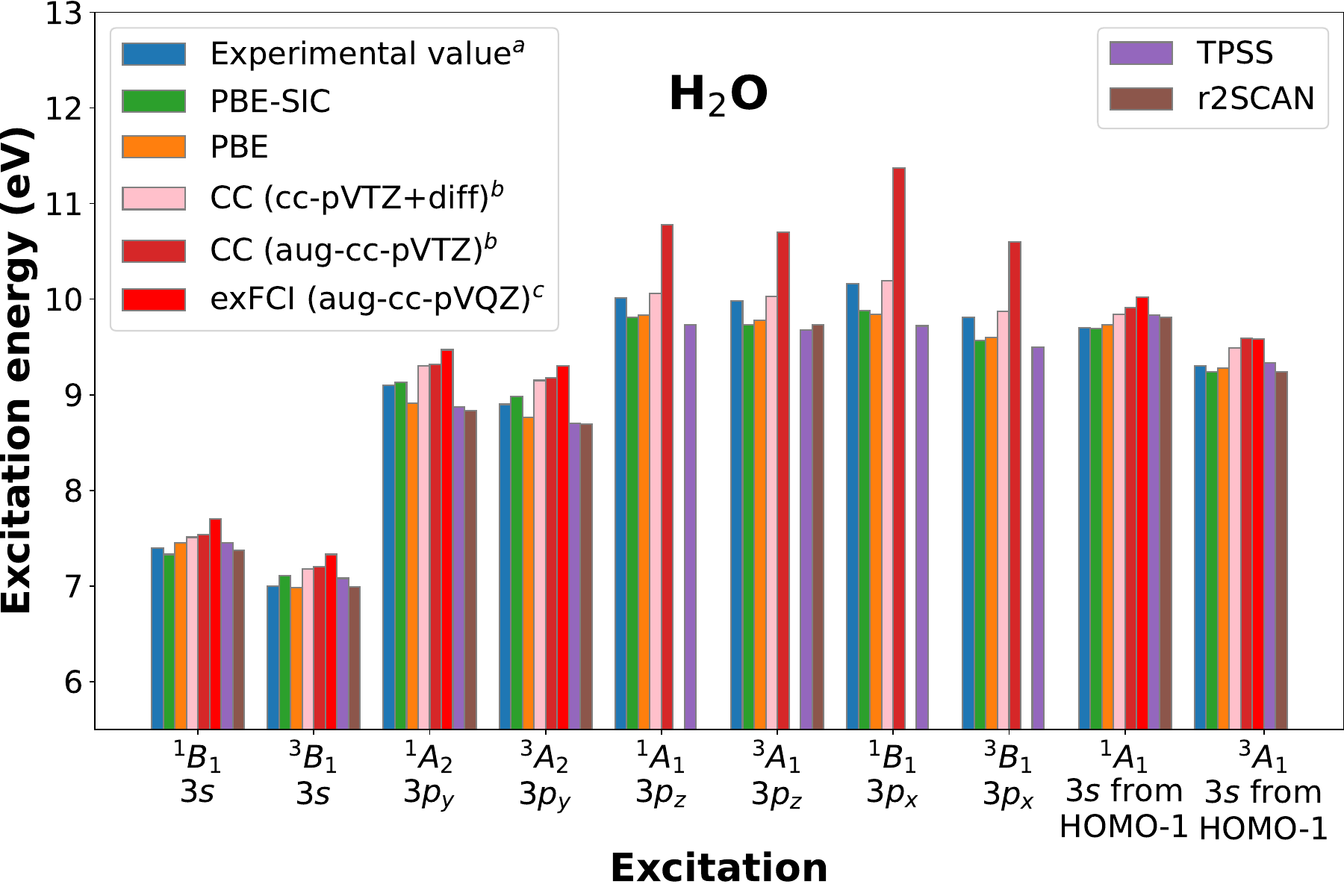}
    \caption{
    Vertical excitation energy for Rydberg states of a water molecule estimated experimentally$^{a,}$\cite{Chutjian1975} (blue), and calculated with PBE-SIC (green) and PBE (orange) functionals. 
    Published results of coupled cluster calculations$^{b,}$\cite{Li2006} with a cc-pVTZ+diff basis set (pink) and aug-cc-pVTZ basis set (dark red) as well as configuration interaction calculations with aug-cc-pVQZ basis set$^{c,}$\cite{Loos2018} (light red), are shown for comparison.
    The cc-pVTZ+diff basis has more diffuse functions than the aug-cc basis sets, which lowers the calculated excitation energy significantly, illustrating the confinement effect of limited basis sets in LCAO calculations, as also illustrated in figure 3.
    Results of calculations with meta-GGA functionals, TPSS (purple) and r2SCAN (brown) are also shown and turn out to be quite similar.  
    }
    \label{fig:water}
\end{figure}

The results for the water molecule are shown in figure~\ref{fig:water}. The PBE-SIC calculations give excitation energy values close to the experimental estimates,\cite{Chutjian1975} except for the 3$p_x$ and 3$p_z$ states
where a lower energy is obtained. 
Results of quantum chemistry calculations using a coupled cluster method and several LCAO basis sets have been published by Li and Paldus.\cite{Li2006} They show that a typical aug-cc-pVTZ basis set gives significantly larger excitation energy for the 3$p_x$ and 3$p_z$ states than a cc-pVTZ+diff
basis set which has additional diffuse function. 
Published results of configuration interaction calculations are also shown.\cite{Loos2018}
Results of calculations using the two meta-GGA functionals, TPSS and r2SCAN, are also shown and are again found to be quite similar.

A summary of the comparison between experimental estimates and the results obtained with the PBE-SIC, PBE and TPSS functionals is illustrated in figure 8, where the mean difference and the range of differences for all 31 excitations is shown. The PBE-SIC results are remarkably close to the experimental estimates, with an average difference of only 0.02\,eV for the singlet states and even smaller for the triplet states, 
while the mean absolute error is 0.1\,eV. For the PBE and TPSS functionals, the excitation energy is systematically predicted to be lower than experimental estimates, but the mean absolute error is still relatively small, 0.2\,eV. The r2SCAN results are not included in this summary because convergence could only be reached for 60\% of the states, but from figures 4-7 showing the results for each molecule it is clear that r2SCAN gives nearly the same estimates of the excitation energy as the TPSS functional.  The numerical values of the differences are given in Tables I and II. The numerical results for each one of the states are given in the Appendix.

\begin{figure}
    \centering
    \includegraphics[width=0.7\textwidth]{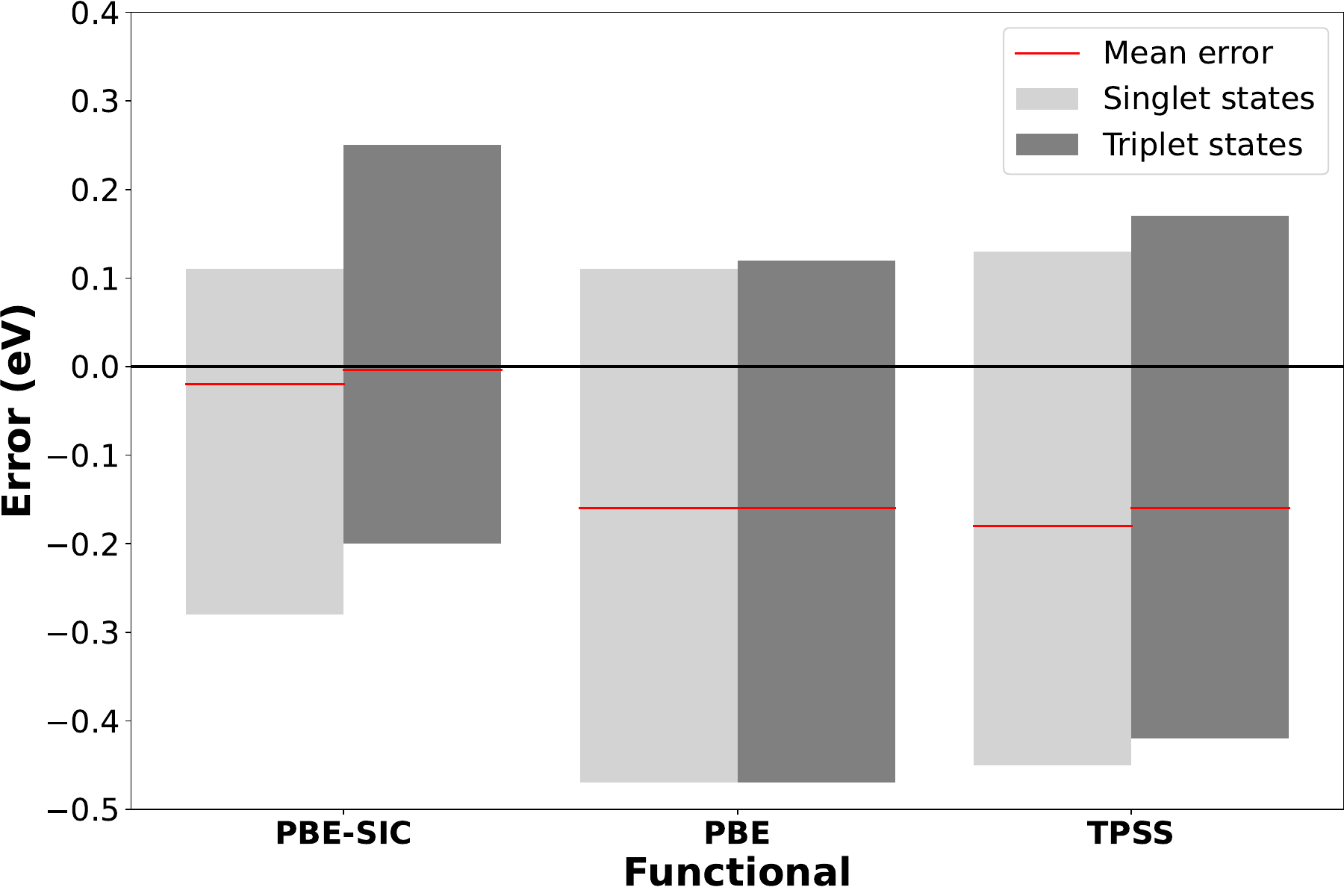}
    \caption{The range of differences between calculated vertical excitation energy using PBE-SIC, PBE and TPSS functionals as compared with experimental estimates. The singlet states are shown in light gray and the triplet states in dark gray. The mean error is shown for each functional with a red line.
    The results obtained with the r2SCAN functional are similar to the ones obtained with TPSS, but are not shown since only 60\% of the excited state calculations could be made to converge with r2SCAN.}
    \label{fig:errors}
\end{figure}

The importance of using complex-valued orbitals in PZ-SIC calculations has been noted previously in studies of atoms and molecules.\cite{Klupfel2011,Klupfel2012,Lehtola2016,Lehtola2016b,Lehtola2014}  Complex-valued orbitals turn out to be important also in the Rydberg state calculations presented here. If the PZ-SIC calculations are carried out with real-valued orbitals, the excitation energy turns out to be systematically lower, even lower than with uncorrected PBE, with average difference from the experimental estimates of 0.4\,eV.


\section{\label{sec:Discussion} Discussion}

Overall, it is remarkable how well the density functionals do in these calculations of high-energy excited states, even though they were developed for ground state calculations. 
The energy of a virtual orbital associated with a Rydberg excited electron is typically positive in ground state PBE calculations, thereby corresponding to an unbound electron, except for the lowest Rydberg state. Nevertheless, the variational calculation of the excited state where the orbitals are optimized for that state gives a bound Rydberg electron and an excitation energy that is quite close to the experimental estimate.
However, the excitation energy obtained with PBE, as well as the more elaborate meta-GGA functionals TPSS and r2SCAN, is systematically lower than the experimental estimate. The use of PZ-SIC in combination with complex-valued orbitals increases on average the calculated excitation energy and gives values that are remarkably close to the experimental estimates, with a mean error of less than 0.02\,eV, and a mean absolute error of less than 0.10\,eV for the four molecules and 31 excited states studied here.
This improvement is likely due to the correction to the long-range form of the effective potential for an electron, as it provides the proper -1/r dependence, while the approximate KS functionals suffer from positive self-interaction even at long range.
While the experimental estimates may not be reliable as values for the vertical excitation energy because of zero-point energy and vibrational energy effects, the comparison with results of quantum chemistry calculations based on coupled cluster and/or configuration interaction methods indicates that the larger excitation values obtained with PZ-SIC are more accurate than the values obtained with the uncorrected KS functionals.

PZ-SIC with real-valued orbitals does not give this improvement, the underestimation being on average 0.4\,eV, even greater than for the uncorrected KS orbitals.
The use of complex-valued orbitals with PZ-SIC allows the calculation to converge on a solution that is lower in energy, both in the ground state and in the excited state, 
but the lowering of the ground state is in general larger than the lowering of the excited state, resulting in larger excitation energy values. 
The use of complex-valued orbitals is particularly important when an electron is placed in one of a set of degenerate orbitals. Then, the multi-determinant character can to some extent be taken into account by breaking the complex conjugation and time reversal symmetries, while avoiding the breaking of the spatial symmetry. 
In the PBE calculations of excitations of the NH$_3$ molecule to a degenerate pair of $p$ Rydberg orbitals, the results obtained using complex-valued orbitals give excitation energy 0.1\,eV greater than calculations with real-valued orbitals.
%

The meta-GGA functionals considered here, TPSS and r2SCAN, do not provide a systematic improvement of the Rydberg excitations obtained with the PBE functional. PZ-SIC turns out to be more successful in this respect than the introduction of higher derivatives of the electron density. The r2SCAN calculations proved to be problematic, as convergence could not be obtained using the grid representation of the orbitals. Instead, a plane wave basis is used in the r2SCAN calculations and even then, convergence has only been obtained for 60\% of the excited states. The computational effort required in the r2SCAN calculations that have succeeded is similar or a bit greater than for the PBE-SIC calculations.

Another important aspect of the results presented here is the efficiency of the real space grid representation of the Rydberg orbitals, where each point in space is equally well represented, as compared to commonly used LCAO basis sets.
Lack of diffuse functions leads to confinement of the Rydberg orbital and thereby, higher excitation energy.
This confinement is evident from the PBE calculations shown in figures 2 and 3 where results of grid and LCAO calculations are compared.
The LCAO basis set is similar to what is used in some of the high-level quantum chemistry calculations. The difference is systematic in that the LCAO approach 
gives higher excitation energy than the real space grid calculations, presumably because the Rydberg state orbital is too confined by the LCAO basis.  The atomic basis can be expanded with extra-diffuse functions 
%
%
and the calculated excitation energy thereby reduced, as showed by Li and Paldus for the H$_2$O molecule,\cite{Li2006} but the mathematical overcompleteness of atomic basis sets can lead to numerical problems if many diffuse functions are added.\cite{Lehtola2019}

While the variational density functional calculations of Rydberg states presented here show promising results, especially when using the PZ-SIC in combination with complex-valued orbitals, more calculations need to be done for a wider range of molecules to further test the accuracy of this approach.  It will be particularly interesting to apply the method to large systems since the computational effort scales relatively slowly with system size.


\section{\label{sec:Conclusions} Conclusions}

Time-independent, variational density functional calculations are found to provide good estimates of Rydberg excitation energy values for several molecules (ethylene, formaldehyde, ammonia and water molecule) when the orbitals are optimized for the excited state, even with approximate KS functionals such as PBE, TPSS and r2SCAN, that have been developed for ground state calculations. The variational calculation can be carried out with a computational effort that is similar to a ground state calculation by converging on saddle points on the electronic energy surface.
The use of Perdew-Zunger self-interaction correction in combination with complex-valued orbitals improves the results and gives remarkably close agreement with experimental estimates, especially for high-energy excitations.  

More testing needs to be done using this approach to see whether it gives good results for other molecules and excitations and to elucidate what features affect the accuracy of the results. 
An interesting future task would be to test the accuracy of SIC applied to meta-GGA functionals. 


\section*{Acknowledgements}
This work was funded in part by the Icelandic Research Fund (grants 217751 and 217734) and by the University of Iceland Research Fund. 
HJ thanks Peter M. Weber for stimulating discussions.
The calculations were carried out at the the Icelandic High Performance Computing Center (IHPC) facility. 


\section*{Data Availability Statement}

%
The data supporting the findings of this study are available within the article and its supplementary material at Zenodo\cite{sigurdarson_2023_10041836}.



\begin{table}[]
\centering
\caption{\label{tab:singlet} Summary of the mean error (ME), mean absolute error (MAE) and root mean squared error (RMSE) of the excitation energy values of the singlet excitations in eV.}
\begin{ruledtabular}
\begin{tabular}{rrrlr}
\multicolumn{1}{l}{Molecule}       & Number of states & PBE-SIC     & PBE       & TPSS  \\ \hline
\multicolumn{1}{l}{C$_2$H$_4$}     & 5                &             &           &       \\
ME                                 &                  & 0.02        & -0.15     & -0.19 \\
MAE                                &                  & 0.05        & 0.16      & 0.20  \\
RMSE                               &                  & 0.06        & 0.18      & 0.22  \\
\multicolumn{1}{l}{H$_2$CO}        & 4                &             &           &       \\
ME                                 &                  & -0.06       & -0.39     & -0.32 \\
MAE                                &                  & 0.06        & 0.31      & 0.32  \\
RMSE                               &                  & 0.07        & 0.36      & 0.34  \\
\multicolumn{1}{l}{H$_2$O}         & 5                &             &           &       \\
ME                                 &                  & -0.07       & -0.12     & -0.18 \\
MAE                                &                  & 0.12        & 0.15      & 0.23  \\
RMSE                               &                  & 0.16        & 0.19      & 0.26  \\
\multicolumn{1}{l}{NH$_3$}         & 3                &             &           &       \\
ME                                 &                  & -0.01       & 0.04      & -0.00 \\
MAE                                &                  & 0.10        & 0.06      & 0.10  \\
RMSE                               &                  & 0.11        & 0.07      & 0.11  \\
\multicolumn{1}{l}{\textbf{Total}} & 17               &             &           &       \\
ME                                 &                  & -0.03       & -0.16     & -0.18 \\
MAE                                &                  & 0.08        & 0.19      & 0.22  \\
RMSE                               &                  & 0.11        & 0.24      & 0.25  \end{tabular}
\end{ruledtabular}
\end{table}


\begin{table}[]
\centering
\caption{\label{tab:triplet} Summary of the mean error (ME), mean absolute error (MAE) and root mean squared error (RMSE) of the excitation energy values of the triplet excitations in eV.}
\begin{ruledtabular}
\begin{tabular}{rrrlr}
\multicolumn{1}{l}{Molecule}       & Number of states & PBE-SIC        & PBE       & TPSS  \\ \hline
\multicolumn{1}{l}{C$_2$H$_4$}     & 5                &                &           &       \\
ME                                 &                  & -0.01          & -0.15     & -0.19 \\
MAE                                &                  & 0.06           & 0.17      & 0.21  \\
RMSE                               &                  & 0.07           & 0.18      & 0.23  \\
\multicolumn{1}{l}{H$_2$CO}        & 3                &                &           &       \\
ME                                 &                  & 0.11           & -0.34     & -0.22 \\
MAE                                &                  & 0.11           & 0.34      & 0.22  \\
RMSE                               &                  & 0.15           & 0.36      & 0.27  \\
\multicolumn{1}{l}{H$_2$O}         & 5                &                &           &       \\
ME                                 &                  & -0.07          & -0.12     & -0.14 \\
MAE                                &                  & 0.15           & 0.12      & 0.19  \\
RMSE                               &                  & 0.17           & 0.15      & 0.22  \\
\multicolumn{1}{l}{NH$_3$}         & 1                &                &           &       \\
ME                                 &                  & 0.02           & 0.12      & 0.17  \\
MAE                                &                  & 0.02           & 0.12      & 0.17  \\
RMSE                               &                  & 0.02           & 0.12      & 0.17  \\
\multicolumn{1}{l}{\textbf{Total}} & 14               &                &           &       \\
ME                                 &                  & -0.00          & -0.16     & -0.16 \\
MAE                                &                  & 0.10           & 0.17      & 0.20  \\
RMSE                               &                  & 0.13           & 0.22      & 0.23  \end{tabular}
\end{ruledtabular}
\end{table}

\clearpage

\vfill\eject

\appendix

\section{Tables}

\begin{table}[h]
\caption{\label{tab:ethylene} The vertical excitation energy (in eV) calculated for various Rydberg states of C$_2$H$_4$ using different functionals.}
\begin{ruledtabular}
\begin{tabular}{lrrrrrr}
\textbf{Excitation} & \textbf{Exp.}\cite{Robin1985} & \textbf{PBE} & \textbf{PBE} & \textbf{PBE-SIC}  & \textbf{TPSS} & \textbf{r2SCAN} \\ 
 & & (aug-cc-pVDZ+sz) & & & & \\ \hline
$^1B_{3u}$ ($\pi \rightarrow 3s$)         & 7.11 & 7.18 & 7.15 & 7.22  & 7.13 & 7.10 \\
$^3B_{3u}$ ($\pi \rightarrow 3s$)         & 6.98 & 7.05 & 7.02 & 7.08  & 7.02 & 7.01 \\
$^1B_{1g}$ ($\pi \rightarrow 3p_y$)       & 7.80 & 7.64 & 7.60 & 7.82  & 7.59 &      \\
$^3B_{1g}$ ($\pi \rightarrow 3p_y$)       & 7.79 & 7.59 & 7.55 & 7.76  & 7.52 &      \\
$^1B_{2g}$ ($\pi \rightarrow 3p_z$)       & 7.90 & 7.79 & 7.74 & 7.91  & 7.70 & 7.77 \\
$^3B_{2g}$ ($\pi \rightarrow 3p_z$)       & 7.79 & 7.66 & 7.62 & 7.81  & 7.58 & 7.60 \\
$^1A_g$ ($\pi \rightarrow 3p_x$)          & 8.28 & 8.53 & 8.07 & 8.22  & 7.96 & 7.92 \\
$^3A_g$ ($\pi \rightarrow 3p_x$)          & 8.15 & 8.31 & 7.94 & 8.03  & 7.86 &      \\
$^1B_{3u}$ ($\pi \rightarrow 3d\sigma$)   & 8.62 & 8.96 & 8.66 & 8.65  & 8.37 &      \\
$^3B_{3u}$ ($\pi \rightarrow 3d\sigma$)   & 8.57 & 8.91 & 8.64 & 8.56  & 8.33 &      \\\end{tabular}
\end{ruledtabular}
\end{table}

\begin{table}[h]
\caption{\label{tab:formaldehyde} The vertical excitation energy (in eV) calculated for various Rydberg states of CH$_2$O using different functionals.}
\begin{ruledtabular}
\begin{tabular}{lrrrrrr}
\textbf{Excitation}                        & \textbf{Exp.}\cite{Clouthier1983,Taylor1982}  & \textbf{PBE} & \textbf{PBE}   & \textbf{PBE-SIC}  & \textbf{TPSS} & \textbf{r2SCAN} \\
 & & (aug-cc-pVDZ+sz) & & & & \\ \hline
$^1B_2$($n \rightarrow 3s$)    & 7.09 & 6.88        & 6.85 & 7.12    & 6.94 & 6.92 \\
$^3B_2$($n \rightarrow 3s$)    & 6.83 & 6.66        & 6.63 & 7.08    & 6.77 & 6.69 \\
$^1B_2$ ($n \rightarrow 3p_z$) & 7.97 & 7.70        & 7.58 & 7.93    & 7.67 & 7.68 \\
$^3B_2$ ($n \rightarrow 3p_z$) & 7.79 & 7.54        & 7.42 & 7.82    & 7.54 & 7.56 \\
$^1A_1$ ($n \rightarrow 3p_y$) & 8.12 & 7.73        & 7.65 & 8.02    & 7.70 &      \\
$^3A_1$ ($n \rightarrow 3p_y$) & 7.96 & 7.56        & 7.50 & 8.01    & 7.60 &      \\
$^1A_2$ ($n \rightarrow 3p_x$) & 8.38 & 8.21        & 7.93 & 8.27    & 7.96 &      \\
$^3A_2$ ($n \rightarrow 3p_x$) &      & 8.22        & 7.93 & 8.32    & 7.97 &      \\\end{tabular}
\end{ruledtabular}
\end{table}

\begin{table}[h]
\caption{\label{tab:ammonia} The vertical excitation energy (in eV) calculated for various Rydberg states of NH$_3$ using different functionals. For the $^1E$ and $^3E$ states, the excitation value obtained without breaking complex conjugation symmetry is shown in parentheses.}
\begin{ruledtabular}
\begin{tabular}{lrrrrrr}
\textbf{State} & \textbf{Exp.}\cite{Skerbele1965,Arfa_Tronc_1991} & \textbf{PBE} & \textbf{PBE} & \textbf{PBE-SIC}  & \textbf{TPSS} & \textbf{r2SCAN} \\ 
 & & (aug-cc-pVDZ+sz) & & & & \\ \hline
$^1A_2$ ($n \rightarrow 3s$)     & 6.38 & 6.45        & 6.42        & 6.35  & 6.49 & 6.39 \\
$^3A_2$ ($n \rightarrow 3s$)     & 6.02 & 6.17        & 6.14        & 6.04  & 6.19 & 6.08 \\
$^1E$ ($n \rightarrow 3p_{\pm}$) & 7.90 & 7.90 (7.84) & 7.88 (7.75) & 7.76  & 7.74 & 7.73 \\
$^3E$ ($n \rightarrow 3p{_\pm}$) &      & 7.72 (7.69) & 7.61 (7.59) & 7.58  & 7.57 & 7.56 \\
$^1A_1$ ($n \rightarrow 3p_z$)   & 8.14 & 9.20        & 8.00        & 8.27  & 8.18 &      \\
$^3A_1$ ($n \rightarrow 3p_z$)   &      & 8.70        & 8.25        & 7.90  & 7.94 &      \\
\end{tabular}
\end{ruledtabular}
\end{table}

\begin{table}[h]
\caption{\label{tab:water} The vertical excitation energy (in eV) calculated for various Rydberg states of H$_2$O using different functionals.}
\begin{ruledtabular}
\begin{tabular}{lrrrrrrr}
\textbf{State} & \textbf{Exp.}\cite{Chutjian1975} & \textbf{PBE} & \textbf{PBE} & \textbf{PBE} & \textbf{PBE-SIC}  & \textbf{TPSS} & \textbf{r2SCAN} \\
 & & (aug-cc-pVDZ+sz) & (cc-pVDZ+diff+sz) & & & & \\ \hline
$^1B_1$ ($n \rightarrow 3s$)   & 7.40  & 7.48  & 7.50  & 7.45 & 7.33  & 7.46 & 7.37 \\
$^3B_1$ ($n \rightarrow 3s$)   & 7.00  & 7.12  & 7.03  & 6.98 & 7.11  & 7.08 & 6.99 \\
$^1A_2$ ($n \rightarrow 3p_y$) & 9.10  & 8.97  & 8.94  & 8.91 & 9.13  & 8.87 & 8.83 \\
$^3A_2$ ($n \rightarrow 3p_y$) & 8.90  & 8.81  & 8.78  & 8.76 & 8.98  & 8.70 & 8.69 \\
$^1A_1$ ($n \rightarrow 3p_z$) & 10.16 & 10.91 & 9.97  & 9.84 & 9.88  & 9.73 &      \\
$^3A_1$ ($n \rightarrow 3p_z$) & 9.81  & 10.77 & 9.74  & 9.60 & 9.57  & 9.50 & 9.73 \\
$^1B_1$ ($n \rightarrow 3p_x$) & 10.01 & 11.24 & 10.07 & 9.83 & 9.81  & 9.72 &      \\
$^3B_1$ ($n \rightarrow 3p_x$) & 9.98  & 10.68 & 10.01 & 9.77 & 9.73  & 9.67 &      \\
$^1A_1$ ($n \rightarrow 3s$)   & 9.70  & 9.80  &       & 9.73 & 9.69  & 9.83 & 9.81 \\
$^3A_1$ ($n \rightarrow 3s$)   & 9.30  & 9.35  &       & 9.28 & 9.24  & 9.33 & 9.24
\end{tabular}
\end{ruledtabular}
\end{table}



\clearpage
 
\bibliography{main}

\end{document}